\font \big =cmbx12 at 20pt
{\big \noindent  What are quantum theorists doing at a \hfil \break
 \noindent  conference on consciousness?}\footnote {$^1$}{Text of talk to be
 given at the Tucson II conference, Towards a Science of Consciousness,
 1996}
 \vskip 1cm
 \centerline {{\it Euan J. Squires}\footnote {$^2$}{e.j.squires@durham.ac.uk}}
             \smallskip        {\it
\centerline {Department of mathematical sciences}
\centerline {University of Durham}
\centerline {Durham City, DH1 3LE, England.}}
\smallskip
\centerline{ February, 1996}

\vskip 1.5cm
{\narrower \smallskip
\noindent {\bf Abstract} The reason why orthodox quantum theory
necessarily invokes consciousness is explained. Several procedures
whereby the Born probability rule can be introduced  are discussed, and
reasons are given for prefering one in which consciousness selects a
unique realised world. Consciousness is something outside of the laws of
physics (quantum mechanics), but it has a real effect upon the
experienced world. Finally,
 orthodox quantum theory is shown to require that consciousness acts
non-locally.\smallskip }
\vskip 1cm

A possible answer to the question in the title of
 this talk would be to say that we can explain
consciousness. That would be false; we have no more idea how to explain
consciousness than anyone else, which means we have NO idea!  In fact
the whole idea of trying to {\it explain} consciousness is probably a
mistake; consciousness just {\it is}.

The proper answer to the question
 is that we cannot understand quantum theory without
invoking consciousness.\footnote {$^3$}{The  interpretation
 problem of quantum theory and the
  hard problem of consciousness also share the property of attracting
  articles that claim to provide a solution but in fact do not address
  the problem.}

Quantum theory is a wonderful elegant theory, which, at least in
principle,  allows us to calculate the properties of all physical (and
chemical) systems. It gives correct results for the probabilities of
particular results of an enormous range of experiments. It is accurate
and universal, and no violations of its predictions are known, even
 where those predictions are very counter-intuitive.

BUT, if quantum theory really applies to all systems, then, except in
very special circumstances, there can never be any
observations, i.e., there can be no events for which the above
statistical predictions could apply.

This fact is so important, and so simple, that we shall discuss
 an example. We
imagine that there is a system, the one to be ``measured'', which is in a
state that I describe as $|+>$. (There is no need to worry what this
means or why I use such a funny notation).

 This
system in measured with  an
 apparatus $A$, and I suppose that after the measurement
the combined system can be described by $|+, A^+>$. What this means
 is that the $+$ reading on the apparatus corresponds to the system
 being  in
 the state $+$. Generally the
apparatus is not isolated but in interaction with an environment (air,
microwave background, black-holes, strings, etc.). Call all this $E$,
then the full state is $|+, A^+, E^+>$. Of course this will in general
be changing with time, but I do not need to indicate this explicitly.

Next, I suppose that Melinda looks at the apparatus.
 This puts her brain, denoted
by $Me$, into a certain state. Hence the  state of the relevant system
is now $$|+, A^+,
E^+, Me^+>.\eqno (1)$$ Everything up to here is perfectly straightforward.
Assuming that I knew the physical structure of the apparatus (not too
difficult to imagine), of the environment (a bit harder) and of Melinda's
brain
(harder still), then the evolution to this final state could be
calculated from the Schr\"odinger equation.

 Notice that it is totally
 irrelevant to the discussion what is the precise nature of the physical
 brain; we need make no assuptions whatever about that, except that the
 brain is a
 physical thing and is therefore described by the laws of quantum theory.

Now we need to introduce the
 bit that we none of us know anything about. Presumably
the $Me^+$ state of Melinda's brain
 corresponds to some sort of ``pattern'' which
her consciousness, knowing something about the apparatus $A$,
 interprets as meaning that the system was in the state
$+$. To be concrete, suppose the pattern is as given in fig. 1.

We can then repeat the above discussion with some other state of the system.
Let us call this state $|->$.
In an obvious generalisation of the notation used above, the state after
measurement will now be $$|-, A^-, E^-, Me^->.\eqno (2)$$ Again there will be some
pattern in Melinda's brain which her
 consciousness recognises as meaning that the
system was in the state $-$. Suppose this time the pattern is as in fig.
2.

At this stage we do not have anything that is different from classical
physics. The calculation yields a particular result which is interpreted
by consciousness. These however are the ``special circumstances''
referred to above, where there is no measurement problem.

The problem arises because it is possible, and indeed in
some cases very easy, to prepare a state which is not $+$ or $-$, but is
a ``bit of each''. We write this as $a|+>\> +\> b|->$, where the relative
magnitudes of $a$ and $b$ tell us something about how much of $+$
compared to $-$ the state contains. In practice most states that are
observed will have this form. (To prevent any possible
 misunderstanding here it is important to emphasise that this new state
 is not just a way of saying that the state might be $+$ or it might be
 $-$, and that we do not know which. Quantum theory allows us to
 discuss this situation - the state is then called a mixed state -
  but it is not what we have here.)

  What happens when we measure this new
state using the same apparatus as before? Here is another piece of
quantum magic. Given the results of the previous calculations, it is a
trivial matter to calculate what happens. In fact the final state
becomes $$a|+, A^+, E^+, Me^+> + \>b|-, A^-, E^-, Me^->.\eqno (3)$$
We do not get one pattern, or the other, but something that is a bit of
one and a bit of the other. There is nothing obviously remarkable about
that; we started with the system in a state which we wrote as a ``sum''
of  system states, and we finish with a similar type of sum
of observed states.

\smallskip
\noindent Now come the surprises.
                                                         \smallskip
  1. It is a simple matter of fact that Melinda  will
experience either the $+$ state or the $-$ state. As far as her
 experience is concerned the state (3) will be {\it exactly} as
 if it were either that in (1) or that in (2).

  To be careful here, I should say that this is what Melinda
will tell us. In order to be sure (see below) we can check by doing the
observation  ourselves. Then I will
 be aware of either the pattern of fig 1 or the pattern of
figure 2 - certainly not the ``sum'' of the two patterns as in fig. 3.

Notice  that this one result of which I am aware exists in Consciousness,
but not in ``physics'', i.e., no particular result is present in the
state of the physical world as it is
 calculated according to quantum theory. If this is ``dualism'' then I
 am happy to be called a  dualist.

                                         \smallskip
 2. The above  fact follows from orthodox quantum theory - in the
sense that will be made clear immediately. Because I sometimes read
statements that seem to deny this, and because it is important to be
precise about what is meant, and because, if we think about it, it is
 rather amazing, I shall derive it.

To do this I suppose that Melinda had  agreed
 to write a $0$ on a piece of paper as soon as she
knew whether the system was in the state $+$ or the state $-$. Note that
she does not write down {\it what} it is, only that she {\it knows} what it is.
  Clearly, in
both the cases where the system was in the state $+$ or $-$, as soon as
she had looked at the apparatus, then she would have written the $0$. It
then follows trivially from standard quantum theory
that in the case where the state was the sum of
the two, she would again write down the $0$. Thus, she would tell the world
that she had become aware of either the pattern in fig. 1, or the pattern
in figure 2. Every physical action she took would convey this message.
 I think this means that she  would indeed have become aware of one
 result, as indeed we know happens in practice.
Otherwise she  would consistently be telling the world, and herself,
 a lie.

 This seems to imply  that
orthodox quantum theory has told us something about how
consciousness actually works. Of course it cannot do this. We have
inserted an assumption of consistency. All Melinda's
  physical actions, as long
as they are governed by quantum theory will imply that she knows a unique
result. The assumption is that this actually means that she does know such a
result. One could imagine (just about) that, on the contrary, she was
 not aware of any
result, but that nevertheless she  put the $0$ on the paper, and in all
other ways behaved as though she did.

It is worth noting that we cannot run this argument with a computer (try
it). It works because Melinda is conscious and it therefore makes sense
to talk about ``knowing''. Computers, on the other hand do not  {\it know}
 anything, and we would not have any way of giving the essential
 instruction to write a $0$ as soon as the result is known.
                                                       \smallskip
 3. Here is something that does not follow from the simple
evolution equation of quantum theory, i.e., the Schr\"odinger equation,
 which is all we have used so far.
In a large set of identical runs  of the above experiment the number of
times Melinda would see $+$
and $-$ would be in the ratio $|a|^2 /|b|^2$. This is a ``rule'' which is
sort of added to quantum theory. It is called the Born rule (after Max
Born who first proposed it), and it has been  confirmed repeatably in
myriads of experiments.
                    \smallskip
So, where is the problem, and what
 has all this got to do with consciousness?

\noindent    { \it The complete
description of the ``physics'' in orthodox quantum theory is the state
displayed  above, which contains both terms, i.e. both ``results''.
 The unique result of which I am aware does not exist in
physics - but only in consciousness. The Born rule does not have
anything to say about physics - it says something about
consciousness.\smallskip }

I must qualify the above by emphasising the fact that I am speaking of
{\it orthodox} quantum theory. I could add something to physics (e.g.
the Bohm hidden variable model) or I could change it (e.g. the explicit
collapse models of GRW/Pearle etc), so that the result would be in the
physics. Even then the properties of consciousness would appear, but all
that is another story which we shall not follow here.
 \vskip 1cm
\noindent {\bf Naive Many-Minds Interpretation}
\smallskip
To continue, we note that the simplest possibility for what is
happening would be that after the measurement there are two ``Melinda's'',
one of which has one experience, and one of which has the other. We need
have no concern that this does not appear to Melinda to be what is happening,
because it is guaranteed by quantum theory that each ``Melinda''
 will be unaware of
the existence of the other, and will indeed have no possibility of knowing
about the other (this is true for all practicable purposes - if she  were
sufficiently clever she could perhaps devise means of checking whether
the other Melinda really exists).
 What we have here is the ``naive'' many-worlds
interpretation of quantum theory; it is better called the ``many-views''
or ``many minds'' interpretation [1,..4] because the physical world, described
by the quantum state, e.g., as displayed above in our simple example
(eq. 3),
 is always ONE thing.

Two points should be noted here. First, the experienced world is
precisely that, {\it the world as experienced}. It is not identical to the
physical world. When we ``measure'' something, we experience a
particular result, but, in general, that result does not refer to
anything that was there before our experience of it, or even after the
experience; it exists only in consciousness. Secondly, all this has been
achieved with nothing beyond orthodox quantum theory.

However, although it is superficially very attractive,
 this naive interpretation  DOES NOT WORK. The reason is simple -
it contains no probabilities, i.e., no Born rule. There are not
``degrees of existence''; everything will exist regardless of how small
its probability should be according to the Born rule. To put this
another way,  probabilities are for something to ``happen'' and here
nothing has actually happened. Now I am aware that the foundations of
the whole theory of proability are very unsure, even in the classical
domain, but this should not prevent us from recognising that
at this stage we do not have a satisfactory theory of the
 quantum world.

\vskip 1cm
\noindent {\bf One Mind Interpretation}
\smallskip
To make progress, we can propose instead that, although the description
of the physics is as given by the state above, with both terms,
consciousness actually {\it selects} one term [
5,6,7].  Normally this will happen
at random with the weights given by $|a|^2$ and $|b|^2$, so that the Born
rule is guaranteed. (In general to say that something happens at random
requires that we give a weight, and there really are no other
possibilities, so the Born rule is very natural.)

What we have now is I believe an acceptable solution to the measurement
problem of quantum theory. It has several merits.

1. In principle it allows for consciousness to be ``efficacious'', i.e.,
to be able to change the experienced world. In other words it can help
to explain what consciousness is for. The point here is that there
 may be circumstances in which there is a quantum superposition in the
 brain which is not correlated to things outside the brain (like in the
 displayed state above). Then the selection, which perhaps need not be
 random, could determine the
  action that a person takes. This would correspond to our experience of
  free-will, and it would have an effect on the
  experienced world, although it would not alter the total wavefunction.
  In other words it would not violate the requirement (for some people)
  that physics is
  `closed''.

Of course, at this stage of our discussion (but not before) we have to
make some assumptions about physical brains. In order for there to be
the possibility of what we are describing here to happen we have to accept
that brains are genuinely {\it quantum} systems that cannot be described
by classical physics. I do not find this difficult. Although surgeons
may see brains as warm wet matter, which from their point of view can be
described perfectly well by classical physics, it remains true that
there is no such thing as classical matter. Without quantum theory there
would be no matter. To say that quantum effects, as we are describing
here, cannot occur in brains, would be rather like telling a nineteenth century
physicist who had just happened
 to have invented  quantum theory that, even if it were
true, there would
be no possibility of ever detecting its effects in the real world.

2. The association of consciousness with ``selection'' seems to be
something that others, from very different arguments, want  to make.
For a recent example Cotterill [8]  writes ``consciousness is {\it
for}....evaluation and choice''.

3. There would be a unique ``real'', i.e., experienced, world. The
non-selected parts of the wavefunction would not really exist.
 To have an analogy here, imagine a sheet of white
paper. By putting a suitable mask on this we could obtain a picture of
say a person -see figure 4. Now different masks would produce different
pictures ( worlds) - indeed all possible pictures. It would be a misuse
of language however to say that the sheet of white paper {\it contained}
 all the
pictures - only the one selected by the mask would exist.

4. As with most versions of the ``many-worlds'' interpretations, this
allows us to use anthropic arguments to explain the apparent
coincidences necessary for our existence, but here it is with a unique
world, rather then a scenario in which all things conceivable actually
exist. The argument would be that in some sort of universal
wavefunction, consciousness selects a part in which consciousness can
exist.
        \vskip 1cm
\noindent {\bf Alternative Many-Views Models.}
\smallskip
Several attempts have been made  to give a meaning to
probability when all experiences occur. For example, Albert and Loewer
[3] and Albert [9]
have suggested that associated with every person there are a large
number of minds, and that each selects at random as with the single
experience proposal above. Again this seems to work, but clearly the
number has to be very large, otherwise there will be the possibility of
meeting ``zombies''. In fact Albert and
 Loewer suggest an infinite number, which I
find hard to accept, because I am not sure that I really know what it
means to have an infinite number of ``objects'' associated with a given
person. Even worse is the fact that they want a continuous infinity.
This runs into the problem that there is no natural measure on a
continuous
infinity: it just does not mean anything to say, for example, that
``more'' minds see one result than another.  The same problem is met by
Lockwood [4, 10] who proposes instead to have all ``minds'' labelled by a
continuous parameter, e.g., $0 < \lambda <1$, so that a certain fraction
of the line goes to one result, and another fraction to another, etc.,
in each
case so as to give the Born rule. Again, this suffers from what seems
to me to be the insuperable problem  that there is no natural
measure on such a line.

I should add that, on aesthetic grounds, I myself am more comfortable
with the idea that there is one world, rather than having to accept that
 all
things that {\it can be} actually {\it are},
 however improbable the Born rule would
make them. It just seems too much to have to believe that there really
are people, holding conferences on physics and consciousness, etc., who
have never experienced interference, or read about it or met anybody who
had! They are going to have an awful shock next time they see a thin
film of oil on water (or at least ``a large part'', whatever that may
mean in this context, of all of them are!)

\vskip 2cm
\noindent {\bf Non-locality}
\smallskip
Finally, we must discuss the issue of non-locality. It is sometimes
stated that one of the advantages of the ``many-worlds'' style of
solutions to the measurement problem is that they do not suffer from the
non-locality which is all too evident in the Bohm model or in collapse
models. To some extent this is true; the non-locality is removed from
the physics because it only arises from the results of measurements, and
so does
not occur if there are no such results. However, it is still around;
 it has simply been removed to ``consciousness''.

We can see this if we consider how consciousness can take note of the
quantum probability. To do this we need to think a little more about how
we locate the ``patterns'' that correspond to a given experience.
Suppose that the quantum state is given by $|\Psi (x, y, t)>$, where $x$
stands for the variables of particles in the brain and $y$ for particles in
the system, the apparatus and the environment. The displayed state (3) above
is just one particular example of such  a state. To see a pattern we
must project this onto a state of some presumed ``consciousness basis''
in the brain. If we denote this by states $|C_n(x)>$, where the $n$
labels possible experiences, then the probablity of the $n^{th}$
experience is, according to quantum theory, $|<C(x)|\Psi (x, y, t)>|^2$.
This however is not a number, but a function of the positions of all the
other particles (some of these may well be thousands of miles away!).
 To get a number we must
integrate over all these positions. This of course is horrendously
non-local in realistic measurement situations. In other words,
consciousness, if it is to ``know about'' probabilities, as it must if we
are to obtain the Born rule, cannot be a local thing.

This  is very important because it means that in the selection model
there will only be one selection, not one for every separate person, a
fact which ensures the essential property
 that all observers will make the same selection. Another way of saying
 this is to say that consciouness must be thought of as being ONE thing.
This is something in which Schr\"odinger firmly believed, and it may be
a contribution that quantum physics can make to the study of
consciousness, thereby guaranteeing that quantum physicists
 continue to have a place at
meetings like this.

\vskip 1cm
\noindent {\bf Related Ideas}
\smallskip
The idea that consciousness has to be introduced in order to understand
quantum theory has been around since the 1920's.
Apart from the work mentioned above, recent contributions are due
to
Hodgson [11],  Mavromatos and Nanopolous
[12], Penrose [13], Page [14] and  Stapp [15]. These models have many
features in common, and in common with the model that I am advocating
here (in particular, the selection model shares many of the features
discussed by Stapp in his recent articles [16,17]).
The principle difference is that, in varying ways these
authors have models in which the operation of consciousness is
associated with some sort of explicit wavefunction collapse, so that the
physics is not given exactly by the Schr\"odinger equation.
It seems to follow that there will be observable differences between
the predictions of these models and those of standard quantum theory
(cf., for example, [18]). This is not necessarily a bad thing, but the
models need to be made sufficiently precise in order that these
differences can be calculated.

There may be a more serious objection in that a proper description of
the collapse  requires a new equation to replace the
Schr\"odinger equation. Examples of such equations already exist of
course (see [19] and references therein), but, at least in the context of
the present discussion, they suffer from the fact
that there is no reason why the collapse effect has anything to do with
consciousness.\footnote {$^4$}{Some people would regard this as a virtue
 of these models, but they would be unlikely to  attend
  this conference.} Rather the collapse is a universal phenomena, with the rate
being very small, i.e. negligible,
 for microscopic systems,  but being proportional to something like the
 number of particles, so that it is large in the macroscopic world. If we
 follow this line too closely then we are in danger of saying that
 consciousness arises, like rapid collapse, simply from having large
 systems. I believe Stapp, at least, would reject this suggestion, in my
 opinion rightly. Maybe things look different if the stochastic nature
 of the collapse process arises from something that is non-computable.
 This might provide a link with possible non-algorithmic aspects of
 conscious thought.

\vskip 1cm
\noindent {\bf REFERENCES}
\smallskip \noindent 1.
 H.D.Zeh, 1981, The problem of conscious  observation in
quantum mechanical description, {\it Epistemological letters} {\bf 73}
\hfil \break \noindent 2. E.J.Squires, 1987, Many views of one world - an
interpretation of quantum theory, {\it Eur. J. Phys.,}{\bf 8}, 171-174
\hfil \break
\noindent 3. D.Albert and B.Loewer, 1988, Interpreting the many-worlds
interpretation, {\it Synthese}, {\bf 77}, 195-213\hfil \break
\noindent 4. M.Lockwood, {\it Mind, Brain and the Quantum} (Blackwell,
Oxford, 1989)\hfil \break
\noindent 5.  E.J.Squires, {\it Conscious Mind in the Physical World}
(IOP, Bristol, 1990)\hfil \break
\noindent 6. E.J.Squires, 1991, One mind or many?, {\it Synthese,} {\bf
89}, 283-286\hfil \break
\noindent 7. E.J.Squires, 1993, Quantum theory and the relation between
conscious mind and the physical world, {\it Synthese,} {\bf 97}, 109-123
\hfil \break
\noindent 8. R.M.J.Cotterill, 1995, On the unity of conscious
experience, {\it J. Consc. Studies}, {\bf 2}, 290-311 \hfil \break
\noindent 9. D.Albert, {\it Quantum Mechanics and
Experience} (Harvard University Press, Cambridge, 1992)\hfil \break
\noindent 10. M.Lockwood, 1996,  Many-minds interpretations of
quantum mechanics, Oxford preprint,
to be published in the {\it British Journal for the
Philosophy of Science}\hfil \break
\noindent 11.  D.Hodgson, {\it The Mind Matters}    \hfil \break
\noindent 12. N.Mavromatos and D.V.Nanopolous, 1995,
Non-critical string theory
formulation of microtubule dynamics and quantum aspects of brain
function, CERN preprint TH/95-127\hfil \break
\noindent 13. R. Penrose, {\it The Emperor's New Mind} (Oxford, 1989)
\hfil \break
\noindent 14. D.Page, 1995, Sensible quantum mechanics: are only
perceptions probabilistic?, Alberta preprint \hfil \break
\noindent 15. H.P.Stapp,
 {\it Mind, Matter and Quantum Mechanics} (Springer-Verlag,
Berlin, 1993)\hfil \break
\noindent 16. H.P.Stapp, 1996, Chance, choice, and consciousness; a causal
quantum theory of the mind/brain, Berkeley preprint, LBL-37944MOD,
invited talk at this conference. \hfil \break
\noindent  17. H.P.Stapp, 1996, The hard problem: a quantum approach,
Berkeley preprint, LBL-37163MOD, to be published in {\it  Journal of
Consciousness Studies}\hfil \break
\noindent 18. P.Pearle and E.J.Squires, 1994, Bound state excitation, nucleon
decay experiments, and models of wavefunction collapse, {\it Phys. Rev.
Letters}, {\bf 73}, 1-5\hfil \break
\noindent 19. G.C.Ghirardi, P.Pearle and A.Rimini, 1990, Markov processes
 in Hilbert space and continuous spontaneous localisation, {\it Phys.
Rev.,} {\bf A42}, 78-89
\vskip 1cm
\noindent {\bf FIGURE CAPTIONS}
\vskip 1cm
\noindent  Fig. 1. A hypothetical neural pattern which Melinda
interprets as the result $+$.
\smallskip
\noindent Fig. 2. A hypothetical neural pattern which Melinda interprets
as $-$.
\smallskip
\noindent Fig. 3. A possible ``sum'' of neural patterns corresponding to
the superposed state. But Melinda's experience corresponds either to the
pattern in fig. 1 or to that in fig. 2.
\smallskip
\noindent Fig. 4. A template that produces a pink man from a sheet of
pink paper. Is the man already present without the template?
\end